\documentstyle[12pt, amsbsy]{article}
\begin{document}

\begin{center}

{\LARGE \bf Critical velocities in exciton superfluidity}
\vspace{10mm}

{\large \bf I.~Loutsenko }

\medskip

{\em Centre de Recherches Math\'ematiques, Universit\' e de Montr\' eal, \par
C.P. 6128, succ. Centre-ville, Montr\' eal, Qu\' ebec, H3C 3J7, Canada 
\par e-mail: loutseni@crm.umontreal.ca }

\vspace{5mm}

{\large \bf D.~Roubtsov }

\medskip

{\em Groupe de Recherche en Physique et Technologie des Couches Minces, \par
D\'epartement de Physique, Universit\' e de Montr\' eal, \par
C.P. 6128, succ. Centre-ville, Montr\' eal, Qu\' ebec, H3C 3J7, Canada 
\par e-mail: roubtsod@physcn.umontreal.ca }

\end{center}

\begin{abstract}

The presence of exciton phonon interactions is shown to play a key role in the exciton
superfluidity. We apply the Landau criterion for an exciton-phonon condensate moving uniformly
at zero temperature. It turns out that there are essentially two critical velocities in the theory.
Within the range of these velocities the condensate can exist
only as a bright soliton. The excitation spectrum and differential
equations for the wave function of this condensate are derived.

 
\end{abstract}

The problem of critical velocities in the theory of superfluidity arose a long
time ago when the experiments with the liquid He showed a substantial discrepancy with quantum-mechanical
predictions.
Later, the effect was analyzed and its phenomenological description was given (e.g. see \cite{0}).
The fact that the liquid He could not be treated as a weakly non-ideal Bose gas was
believed to be the main reason for inconsistency of microscopic theory with experimental data.

For a long time He has been the only substance where the superfluidity can be
observed. The recent experiments with the dilute gas of excitons \cite{01}, \cite{02}
provide new possibilities for studying different types of superfluidity.

In this series of experiments the $ {\rm Cu_2 O } $ crystal was irradiated with laser
light pulses of several ns duration. At low intensities of the laser beam
(low concentration of excitons) the system revealed a typical diffusive behavior of exciton gas.
Once the intensity of the beam exceeds
some value, the majority of particles move together in the packet. Their common propagation
velocity is close to the longitudinal sound velocity, and the packet evolves as a bright soliton.

Some alternative explanations of the phenomena are known. One of them \cite{01} implies
that the bright soliton is a one-dimensional traveling wave which satisfies
the Gross-Pitaevsky (nonlinear Schr\"odinger) equation \cite{00} for the Bose-condensate wave function 
$\Psi({\bf x},t)$
\begin{equation}
i\hbar{\partial\Psi\over\partial t}=-{\hbar^2\over 2m}\Delta\Psi+\nu\Psi^*\Psi^2
\label{NS}
\end{equation}
with attractive potential of exciton-exciton interaction $\nu<0$.

A quantitative treatment given in \cite{03} provides an iterative solution for the Heisenberg
equation with the use of perturbational methods. In this picture the second order interactions,
neglected in the Bogoliubov approximation, contribute to the negative value of $\nu$. However,
the influence of exciton-phonon interactions on the dynamics of the condensed excitons is not treated
\cite{03}.

Another interpretation is based on a classical model \cite{05} where the normal exciton gas
is pushed towards the interior of a sample by the phonon wind emanating from the surface.
Such an explanation has a discrepancy with the experiment, because the signal observed
is one order of magnitude longer than the excitation pulse duration \cite{02}.

In this Letter we give an alternative and, in our opinion, more intrinsic interpretation of these
phenomena. We argue that it is a propagation of a superfluid {\it exciton-phonon condensate}
which is observed experimentally. The presence of exciton-phonon interactions is crucial for
a ``soliton-like superfluidity''. This interaction plays a key role
when the propagation velocity approaches the longitudional sound velocity.

We start with the Hamiltonian of the exciton-phonon system
$$
H=H_{\rm ex}+H_{\rm ph}+H_{\rm int},\,
H_{\rm ex}=-{\hbar^2\over 2m}\int\hat{\Psi}^*({\bf x})\Delta\hat{\Psi}({\bf x}){\rm d}{\bf x}+{1 \over 2}
\int\hat{\Psi}^*({\bf x})\hat{\Psi}^*({\bf y})\,\nu({\bf x}-{\bf y})\,\hat{\Psi}({\bf x})\hat{\Psi}({\bf y})
{\rm d}{\bf x}{\rm d}{\bf y},
$$
\begin{equation}
H_{\rm ph}=\int\left\{{1 \over 2\rho}\hat{\boldsymbol
 \pi}({\bf x})^2
+{c^2\rho \over 2}\left(\nabla\,\hat{\bf u}({\bf x})
\right)^2
\right\} {\rm d}{\bf x}, \quad H_{\rm int}=\int \sigma(x-y)\hat{\Psi}^*({\bf x})\hat{\Psi}({\bf x})\left(\nabla\,
\hat{\bf u}({\bf y})\right) {\rm d}{\bf x}{\rm d}{\bf y},
\label{Hamiltonian1}
\end{equation}
where $\hat{\Psi}$ and $\hat {\bf u}$ are the operators of the exciton and phonon fields correspondingly,
$ c $ is the longitudional sound velocity and $ \rho $ denotes the mass density of the
crystal. The field variables obey the following commutation relations
$$
\left [\hat{\Psi}({\bf x}),\hat{\Psi}^*({\bf y})\right ]=
\hbar\delta({\bf x}-{\bf y}),\quad\left [\hat{\pi}_i({\bf x}),\hat{u}_j({\bf y})\right ]=-i\hbar
\delta_{ij}\delta ({\bf x}-{\bf y}), \, i,j=1,2,3 .
$$
In (\ref{Hamiltonian1}) we omit the terms with the transverse sound velocity, since
the interaction of excitons with transverse sound waves is much weaker than with the longitudional
ones.

It is convenient to change the reference system when we consider a uniform motion of the Bose gas.
The transition to the reference system moving uniformly with the velocity ${\bf v}=(v,0,0)$ is immediate.
In new coordinates the classical field equations become:
\begin{equation}
\left(i\hbar{\partial\over\partial t}+{\hbar^2\over 2m}\Delta+{mv^2\over 2}-
\int\!\!\nu({\bf x}-{\bf y})\vert\psi({\bf y},t)\vert^2{\rm d}y^3\right)\psi({\bf x},t)=
\psi({\bf x},t)\int\sigma({\bf x}-{\bf y})\left(\nabla {\bf u}({\bf y},t)\right){\rm d}{\bf y}
\label{aa}
\end{equation}
\begin{equation}
\left({\partial^2\over\partial t^2}-2v{\partial^2\over\partial t\partial x_1}+
v^2{\partial^2\over\partial x_1^2}-c^2\Delta\right){\bf u}({\bf x},t)=
{1\over\rho}\nabla\int\sigma({\bf x}-{\bf y})\vert\psi({\bf y},t)\vert^2{\rm d}{\bf y},
\label{bb}
\end{equation}
where $\psi({\bf x},t)=\Psi(x_{1}+vt, x_2, x_3,  t)\exp\left(-imvx_{1}/\hbar\right)$.

The l.h.s. of Equation (\ref{aa}) is Galileian invariant, while the l.h.s. of (\ref{bb}) is
Lorentz invariant. As a result, the system (\ref{aa}), (\ref{bb}) is neither Galileian
nor Lorentz invariant.
As we will see later, it is due to this noninvariance that the effective potential of
exciton-exciton interactions depends on velocity.

Let us consider slowly varying solutions of the system (\ref{aa}), (\ref{bb}).
In this (long wavelength) limit one can replace
$\nu({\bf x})$ and $\sigma({\bf x})$ by $\nu_0\delta ({\bf x})$ and $\sigma_0\delta ({\bf x})$,
where $\nu_0\,(>0)$ and $\sigma_0$ 
denote the zero-mode Fourier components of the corresponding potentials.

Solving (\ref{bb}), one can
express the bounded at infinity time-independent solution ${\bf u}(x)$
in terms of $\psi({\bf x})$.
The effective potential of the exciton-exciton interaction is obtained after substituting this expression
into (\ref{aa}).
The phonon field makes this potential 
long-range, anisotropic and $v$-dependent. The potential becomes asymptotically attractive
along the ${\bf v}$-direction and asymptotically repulsive in directions perpendicular to ${\bf v}$.
It follows that stability of the corresponding
solutions $\psi=\phi(x_1)\exp(-i\omega_0 t)$, $u_i=\delta_{i1}q(x_1)$
is preserved under the one-dimensional reduction of the system (\ref{aa}), (\ref{bb}).
The functions $\phi(x_1)$, $q(x_1)$ obey the following equations
\begin{equation}
\left({\hbar^2\over 2m}{\partial^2\over\partial x_1^2}-\lambda\right)\phi(x_1)=
\left(\nu_0-{\sigma_0^2\over(c^2-v^2)\rho}\right)\phi(x_1)^3,
\quad\lambda=-\hbar\omega_0-{mv^2\over2}+ C\sigma_0,
\label{classical1}
\end{equation}
\begin{equation}
{\partial q(x_1)\over\partial x_1}=C-{\sigma_0\phi(x_1)^2\over(c^2-v^2)\rho},
\label{classical2}
\end{equation}
where the integration constant $C$ is fixed by the condition $q\to{\rm const}$ as $\vert
x_1\vert\to\infty$. In the last equations $\phi$ assumed to be real. This choice does not change the result
but simplifies our calculations.

It follows from (\ref{classical1}) that the effective potential becomes attractive when $v$ exceeds
the critical velocity
\begin{equation}
v_0=\sqrt{c^2-\left(\sigma_0^2/\nu_0\rho\right)}.
\label{critical}
\end{equation}
But if $v$ exceeds the sound velocity $c$, the potential becomes repulsive again.
As for the solution varying in the direction ${\bf n}=(n_1,n_2,n_3)$,
$\psi=f({\bf n}\,{\bf x})\exp(-i\omega_0 t)$, ${\bf u}= {\bf n}q({\bf n}\,{\bf x})$, the critical velocity is
$v_0({\bf n})=v_0/\cos(\theta), \vert\cos(\theta)\vert>v_0/c$, where $\theta$ is
the angle between ${\bf n}$ and ${\bf v}$. 

When $v$ is less than the critical velocity (\ref{critical}),
equations (\ref{classical1}),(\ref{classical1}) have the following stable stationary solutions
$$
({\rm i})\quad\phi=\phi_0=\sqrt{N/V}={\rm const},\quad {\bf u}={\rm const},\quad
{\rm and}
$$
\begin{equation}
({\rm ii})\quad\phi=\phi_0\tanh\left(\beta\phi_0(x_1-a)\right),
\quad{\partial q(x_1)\over\partial x_1}=
-{\sigma_0\phi_0^2\over(c^2-v^2)\rho}\cosh^{-2}\left(\beta\phi_0(x_1-a)\right),
\label{under}
\end{equation}
$$
\beta=\sqrt{{m\nu\over \hbar^2}{\vert v_0^2-v^2 \vert\over\vert c^2-v^2\vert}},
\quad\lambda=\left({\sigma_0^2\over(c^2-v^2)\rho}-\nu_0\right)\phi_0^2=
\nu_0\phi_0^2{v^2-v_0^2\over c^2-v^2},
\quad C={\sigma_0\phi_0^2\over(c^2-v^2)\rho}\,.
$$
In (\ref{under},i) $N$ and $V$ stand for the number of particles in the condensate and the volume
of the system. 

When $v$ exceeds $v_0$, we have only one stable stationary solution
$$
\phi=\phi_0\cosh^{-1}\left(\beta\phi_0(x_1-a)\right),
\quad{\partial q(x_1)\over\partial x_1}=
-{\sigma_0\phi_0^2\over(c^2-v^2)\rho}\cosh^{-2}\left(\beta\phi_0(x_1-a)\right)  ,
$$
\begin{equation}
\quad\lambda={\phi_0^2\over 2}\left({\sigma_0^2\over(c^2-v^2)\rho}-\nu_0\right)=
{\nu_0\phi_0^2\over 2}{v^2-v_0^2\over c^2-v^2},
\quad C=0.
\label{over}
\end{equation}
To find the excitation spectrum of the system we expand the
field operators near the proper classical solutions:
$$
\hat{\psi}({\bf x},t)=\left(\phi(x_1)+\hat{\chi}({\bf x},t)\right)e^{-i \omega_0 t},
\quad {\hat u}_i({\bf x},t)=\delta_{i1}q(x_1)+\hat{\eta}_i({\bf x},t).
$$
The Hamiltonian of the system can be written as follows
\begin{equation}
H=H_0+\hbar H_2+\dots  ,
\label{Bogoliubov}
\end{equation}
where $H_0=H(\phi e^{-i\omega_0 t},q)$ stands for the classical part of $H$.
It is important that $H_2$ is bilinear in $\hat{\chi}({\bf x},t),\, \hat{\eta}({\bf x},t)$, whereas
the linear terms are absent
in (\ref{Bogoliubov}) (since the classical fields satisfy
the stationary equations (\ref{classical1}),(\ref{classical2})). From now on we are working
in quasiclassical approximation and neglecting the terms of power greater than one
(in $\hbar$).

The quasiclassical Hamiltonian (\ref{Bogoliubov}) is reduced to the normal form
\begin{equation}
H_2=\sum_i\omega_i \hat{b}_i^* \hat{b}_i + {\rm const},\quad \left [\hat{b}_i,\,\hat{b}_j^*\right ]=\delta_{ij},
\,\left [\hat{b}_i,\hat{b}_j\right ]=0.
\label{normal}
\end{equation}
Indeed, since $H_2$ is a bilinear function of $\hat{\chi},\,\hat{\eta},$
the equations of motion are linear in field operators.
They coincide with the corresponding classical equations
(i.e. equations (\ref{aa}),(\ref{bb}) linearized around
$\psi({\bf x},t)=\phi(x_1)\exp(-i\omega_0 t),\, u_i({\bf x},t)=
\delta_{i1}q(x_1)$):
\begin{equation}
\left(i\hbar{\partial\over\partial t}+{\hbar^2\over 2m} \Delta - \lambda+ C\sigma_0+
\left\{{\sigma_0^2\over(c^2-v^2)\rho}-2\nu_0\right\}\phi(x)^2\right)\chi-\nu_0\phi(x)^2\chi^*-
\sigma_0\phi(x)(\nabla\,{\boldsymbol
\eta})=0,
\label{soliton1}
\end{equation}
\begin{equation}
\left(c^2 \Delta-v^2 {\partial^2\over\partial x_1^2}+
2v{\partial^2\over\partial t\partial x_1}-
{\partial^2\over\partial t^2}\right){\boldsymbol \eta}+{\sigma_0\over\rho}\nabla
\left(\phi(x)(\chi+\chi^*)\right)=0.
\label{soliton2}
\end{equation}
The quantities $\omega_i$ in (\ref{normal}) are characteristic
frequencies of the system (\ref{soliton1}), (\ref{soliton2}).

Let us consider the homogeneous Bose gas moving uniformly with velocity $v<v_0$.
The condensate wave function is given by (\ref{under},i). The differential equations
(\ref{soliton1}),(\ref{soliton2}) have constant coefficients so that the characteristic frequencies
$\omega({\bf k})$ are determined as roots of the following characteristic polynomial
\begin{equation}
(\Omega^2-c^2 k^2)\left[\hbar^2(\Omega+vk_1)^2-{\hbar^2 k^2\over 2m}
\left({\hbar^2 k^2\over 2m}+\left\{2\nu_0-{\sigma_0^2\over(c^2-v^2)\rho}\right\}\phi_0^2\right)\right]-
{\hbar^2 k^2\over 2m}{\sigma_0^2\phi_0^2 k^2\over\rho}=0,
\label{characteristic}
\end{equation}
where $\Omega=\omega({\bf k})-vk_1$ are the excitation frequencies in the crystal reference frame.
In the limit $\sigma_0\to 0$ one gets the Bogoliubov \cite{06} spectrum
$\hbar\omega(k)=\sqrt{{\hbar^2 k^2\over 2m}\left({\hbar^2 k^2\over 2m}+2\nu_0\phi_0^2\right)}$
for the exciton gas
as well as the free phonon spectrum $\Omega=c k$.
When we switch on an exciton-phonon interaction, the spectrum $\omega({\bf k})$ becomes ${\bf v}$-dependent
and the Landau criterion of superfluidity for the homogeneous Bose gas \cite{0} has to be
properly modified.

The transition to the normal state occurs if
there exists such  ${\bf k} \neq 0$ for a given velocity that $\Omega<0$, i.e.
\begin{equation}
\min_{k}(\omega({\bf k})-vk_1)=0.
\label{ll}
\end{equation}
Analyzing (\ref{characteristic}) we obtain the value
of the critical velocity $v_L$ for the homogeneous exciton-phonon gas
$$
v_L=\sqrt{\left(\nu_0\phi_0^2/m\right)\left(1-\left(\sigma_0^2/\nu_0\rho c^2\right)\right)}=
 \left ( \sqrt{\nu_0 N/ mV} /c \right)\,v_0\,.
$$
The quantization near the translationally noninvariant classical solution (\ref{under},ii)
in the region $v<v_0$ yields the same continuous spectrum $\omega({\bf k})$. The only new feature
is that there appears a bounded state at $\omega=0$ in the ${\bf v}$-direction. This fact has
a simple explanation: the family of the solutions (\ref{under},ii) contains an arbitrary
translation parameter $a$, which, in fact, is a collective coordinate. Differentiation
of (\ref{under},ii) with respect to $a$ gives then necessary time independent solution of
(\ref{soliton1}),(\ref{soliton2}). This bounded state
does not affect the quasiclassical excitation spectrum and contributes only to highest approximations
(e.g. see \cite{08}).

If the velocity $v$ exceeds (\ref{critical}), the characteristic polynomial (\ref{characteristic})
has complex roots and there is no stable constant solutions.
The condensate (i.e. classical) wave function turns into the (bright) soliton (\ref{over})
of the one-dimensional nonlinear Schr\"odinger equation (\ref{classical1}).
This solution decreases exponentially.  This allows
us to obtain the continuous spectrum from asymptotics of (\ref{soliton1}), (\ref{soliton2}). We have
$$
\hbar \omega({\bf k})=\lambda+{\hbar^2 k^2\over 2m}
$$
for the exciton branch of the model, and
$$
\omega({\bf k})=ck+vk_1
$$
for the phonon branch. As in the previous case we get a bounded state at zero energy.
We skip the question of existence of other bound states, since it is not essential for our purposes.

The spectrum now has a {\it gap} in the exciton branch which is equal to $\lambda$. In a sense,
the situation is similar to the BCS theory: the exciton-phonon interaction makes the effective
exciton-exciton potential attractive, and the excitation spectrum acquires a gap.

The transition to the ballistic regime is accompanied by the symmetry breakdown:
a new condensate wave function (\ref{over}) is no more translationally invariant.
However, it contains a free translation parameter. We can interpret this as a
phase transition of the second order.

The value $\phi_0$ is readily computed from the normalization condition
$\int \phi({\bf x})^2{\rm d} {\bf x}=N$, and $\lambda$ is then obtained from (\ref{over})
\begin{equation}
\lambda={m\nu_0^2 N^2 \over 8 \hbar^2 S^2}\left({v^2-v_0^2\over c^2-v^2}\right)^2.
\label{lambda}
\end{equation}
In (\ref{lambda}) $S$ denotes the packet cross-section in $x_2x_3$-plane.
When $v$ approaches the longitudional sound velocity $c$, the gap magnitude increases and soliton becomes
more stable. The soliton energy can be estimated from (\ref{Hamiltonian1})
$$
E=N\left\{{m \nu_0^2 N^2\over 24\hbar^2 S^2}{(v^2-v_0^2)\over (c^2-v^2)^3}
\left(v^4+3 v^2 c^2+v_0^2 c^2-5 v_0^2 v^2 \right)+{m v^2\over 2}\right\}+\dots.
$$
It follows from the last formula that $E\to\infty$ as $v\to c$. Roughly speaking,
the soliton effective mass tends to infinity when its speed approaches the
longitudional sound velocity. Then its motion is less subjected to the external forces.

The onset of ballistical regime is determined by the condition $v>v_0$,
while the frictionless propagation of the packet is possible only if the condition
$\min_{k}\left(\omega({\bf k})-vk_1\right)>0$ holds.
The relation between the treshold speed $v_c$ for the frictionless propagation
and the number $N(v_c)$ of excitons in the packet at this speed follows from (\ref{ll})
$$
{v_c (c^2-v_c^2)\over v_c^2-v_0^2}={\nu_0\over 2\hbar}{N(v_c)\over S},
$$
where the function $N(v)/S$ depends on the characteristics of the light source.
As a consequence, $v_c>v_0$ and the velocity $v_c$ is also dependent on these characteristics.

Thus the frictionless soliton motion is observed over the range $v_c<v<c$.
In principle, it is also possible to observe the onset of soliton propagation for some $v$ close
to $v_c$ with $v_0<v<v_c$, but the packet should disintegrate at later stages of the propagation.

It is easy to see that the solution (\ref{over}) is the most stable in the class of
one-dimensional traveling waves moving uniformly with given $v (>v_0)$ and $N$.
We argue that (\ref{over})
is also the most stable solution in the class of all solutions with given $v (>v_0)$ and $N$,
because the effective exciton-exciton potential is attractive in ${\bf v}$-direction and
repulsive in the perpendicular directions. We would like to stress that effective
one-dimensional solutions
of three-dimensional nonlinear Schr\"odinger equations (\ref{NS}) with attractive
potentials do not have the similar properties. In particular, the stability of such solutions
is doubtful \cite{04}.

When the crystal has no impurities, and the boundary friction is absent (the
experiments with $ {\rm Cu}_2 {\rm O} $ are carried out under these conditions,
cf. \cite{01}, \cite{02}), the propagation of solitons is observed at the critical velocity (\ref{critical}).

The corresponding value of $\sigma_0$ is estimated from the deformation potential of $ {\rm Cu_2O } $.
As a result, we obtain $ v_0 \simeq 0.5 - 0.7 c $.
These estimates are in agreement with the experimental data.

To observe the critical velocity $v_c$
the presence of interior friction is important.
Under these conditions, the relations between $v_c$ and the characteristic width of the soliton $l$ is given by
$$
v_c=\hbar/ml .
$$
In the present work we have discussed the properties of the system at zero temperature.
The extension of our results to finite temperatures seems to be a more difficult problem.

We are very grateful to Y.~Berest for valuable remarks
and to L.~Vinet and Y.~L\'epine for stimulating discussions.
We are also grateful to A.Mysyrowicz for drawing our attention
to the work \cite{10} and useful comments.

\end{document}